\documentclass[9pt,twocolumn,twoside]{opticajnl}
\journal{opticajournal} 

\setboolean{shortarticle}{true}
\usepackage{mathrsfs, amsmath, amssymb}

\usepackage{lineno}
\usepackage{xcolor}

\newcommand{\unit}[1]{\ \mathrm{#1}}

\title{Experimental demonstration of frequency downconverted arm length stabilization for a future upgraded gravitational wave detector}

\author[1,*]{Satoshi Tanioka}
\author[1]{Bin Wu}
\author[1]{Stefan W. Ballmer}

\affil[1]{Department of Physics, Syracuse University, Syracuse, New York 13244, USA}

\affil[*]{stanioka@syr.edu}

\begin{abstract}
Ground-based laser interferometric gravitational wave detectors consist of multiple optical cavity systems whose lengths need to be interferometrically controlled.
An arm-length stabilization (ALS) system has played an important role in bringing these interferometers into operational state and enhance their duty cycle.
The sensitivity of these detectors can be improved if the thermal noise of their test mass mirror coatings is reduced.
Crystalline AlGaAs coatings are a promising candidate for this.
However, the current ALS system with frequency-doubled 532 nm light is no longer an option with AlGaAs coatings because the 532 nm light is absorbed by AlGaAs coatings due to the narrow band gap of GaAs.
Therefore, alternative locking schemes must be developed.
In this letter, we describe an experimental demonstration of a novel ALS scheme which is compatible with AlGaAs coatings.
This ALS scheme will enable the use of AlGaAs coatings in current and future terrestrial gravitational-wave detectors.

\end{abstract}

\setboolean{displaycopyright}{false} 

\begin{document}

\maketitle


Ground-based gravitational wave detectors (GWDs) have opened a new window on the Universe by enabling direct detection of gravitational waves (GWs) \cite{Abbott2016, Abbott2021}.
Those GWDs are based on a Michelson interferometer with multiple optical cavities. 
Five length degrees of freedom (DoFs) of these cavities need to be controlled in order to operate the detectors \cite{Aasi2015}.
Their sensing signals are interdependent as the cavities are coupled with each other.
An arm length stabilization (ALS) system is indispensable to robustly bring such an interferometer to the operating state \cite{Mullavey2012, Izumi2012, Staley2014, Akutsu_2020}.
Current terrestrial GWDs, Advanced LIGO, Advanced Virgo, and KAGRA, employ $532\unit{nm}$ wavelength green lasers generated via second harmonic generation (SHG) as an auxiliary laser for the ALS system while the main laser wavelength is $1064\unit{nm}$~\cite{Staley2014,Akutsu_2020, Acernese2023}.
This enables the arm cavity lengths to be pre-stabilized independently from other DoFs, allowing those other DoFs to be locked without interference from the arm cavities. 

The sensitivity of current GWD is partially limited by thermal noise arising from amorphous silica and titania-doped tantala coatings \cite{Gras2018,Buikema2020}.
In order to enhance the GW detection rate, and hence their scientific impact, lower thermal noise mirror coatings need to be installed~\cite{LSC}.
Crystalline gallium arsenide ($\mathrm{GaAs}$) and aluminum-alloyed gallium arsenide ($\mathrm{Al}_{x}\mathrm{Ga}_{1-x}\mathrm{As}$) coatings (referred to as AlGaAs coatings) are a promising coating candidate for future GWDs, resulting in a factor of at least $5$ improvement in coating thermal noise compared to that of current Advanced LIGO~\cite{Gras2018, Cole2023}.
This significant reduction in coating thermal noise will ultimately enhance the astrophysical reach for a binary neutron star merger, 
yielding up to a factor of 40 increase in detection rates \cite{Cole2023}. 

While crystalline AlGaAs coatings can largely reduce thermal noise, they are opaque to the auxiliary $532\unit{nm}$ laser light due to the narrow band gap of GaAs \cite{Adachi1985}.
Therefore, the current ALS with $532\unit{nm}$ green light is not compatible, and a new ALS scheme is required in order to implement AlGaAs coated mirrors.
To solve this problem, a novel ALS scheme was proposed in \cite{Cole2023}. The scheme uses an optical parametric oscillation (OPO) to frequency-downconvert the laser to  $2128\unit{nm}$, eliminating the need for $532\unit{nm}$ green laser light.

In this letter, we describe the experimental demonstration of this ALS scheme.
We developed a tabletop proof-of-concept setup to demonstrate the frequency-downconverted ALS, and evaluated its performance.
For cost effectiveness, the ALS scheme demonstrated in this letter is using dichroic amorphous coatings, custom ordered to match the anticipated interferometer reflectivity requirements, but the scheme can be used with crystalline dichroic AlGaAs coatings with the same reflectivities, as well as any other low thermal noise coating material that excludes visible wavelength due to their band gap structure for instance amorphous silicon coatings \cite{Steinlechner2021}.

\begin{figure}[ht!]
\centering
\includegraphics[width=\linewidth]{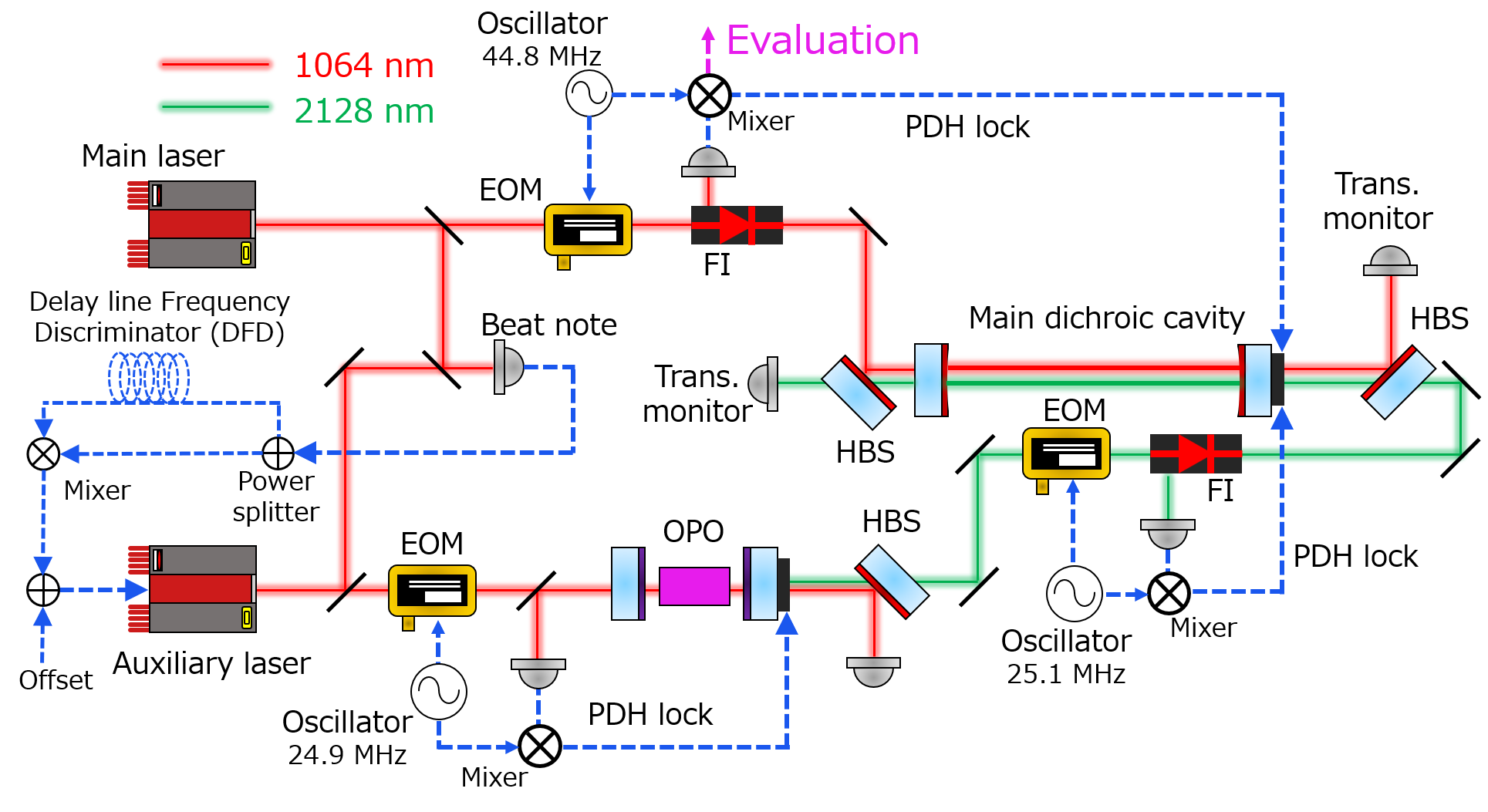}
\caption{Schematic of the ALS demonstration setup. Solid lines are optical paths, and dashed lines are electronic connections.
Two free-running NPRO lasers are used as main and auxiliary lasers.
The main and the auxiliary beams are combined or separated by using a harmonic beam separator.
Reflected beams from the main cavity are picked off by Faraday isolators (FIs).
This setup is installed on an optical table with pneumatic vibration isolators to mitigate seismic vibration coupling.}
\label{Fig:setup}
\end{figure}
\begin{table}[htbp]
\centering
\caption{\bf Main dichroic cavity parameters.}
\begin{tabular}{lc}
\hline
Parameter & Value \\
\hline
Cavity length & $40\unit{cm}$ \\
Free spectral range & $375\unit{MHz}$ \\
Finesse for $1064\unit{nm}$ & $\sim250$ \\
FWHM linewidth for $1064\unit{nm}$ & $2.2\unit{nm}$ \\
Finesse for $2128\unit{nm}$ & $\sim30$ \\
Designed mirror reflectivity & $98.6\%$ at $1064\unit{nm}$ \\
 & $\sim96\%$ at $2128\unit{nm}$ \\
\hline
\end{tabular}
  \label{tab:MainCav}
\end{table}

Fig. \ref{Fig:setup} shows the schematic of the setup. 
We use a $40\unit{cm}$ long Fabry-P\'{e}rot (FP) cavity to represent the main arm cavity.
This FP cavity is composed of a pair of mirrors with amorphous coatings, and both the input and output mirrors have the same reflectivities.
These mirrors are designed to have reflectivities of $98.6\%$ and about $96\%$ at $1064\unit{nm}$ and $2128\unit{nm}$, respectively, to imitate the input test mass (ITM) of Advanced LIGO \cite{Staley2014}.

Both the main and auxiliary $1064\unit{nm}$ laser sources are NPRO lasers made by Coherent.
The beat note is detected by an RF photodiode (RFPD) to measure the frequency difference between the main and the auxiliary lasers.
The delay line frequency discriminator (DFD) is used in our setup in the same manner as previous work \cite{Izumi2012}.
The noise of the delay line is estimated at about $1\unit{Hz/\sqrt{Hz}}$ within our frequency region, i.e., between $0.1\unit{Hz}$ and a few kHz.
The DFD signal is sent to the auxiliary laser piezo transducer (PZT) so that the frequency offset between two lasers is stabilized.
This configuration enables us to keep the main laser off-resonant in the main dichroic cavity, and then bring it to a resonant point by changing the frequency offset.

The OPO cavity is locked to the auxiliary laser by the Pound-Drever-Hall (PDH) method to generate downconverted $2128\unit{nm}$ light \cite{PDH, Darsow-Fromm2020}.
The OPO cavity is composed of a FP cavity and a periodically poled potassium titanyl phosphate (PPKTP) crystal. 
The PDH locking of the OPO cavity is achieved by actuating the cavity length using a PZT attached on the output mirror.
\begin{figure}[ht!]
\centering
\includegraphics[width=\linewidth]{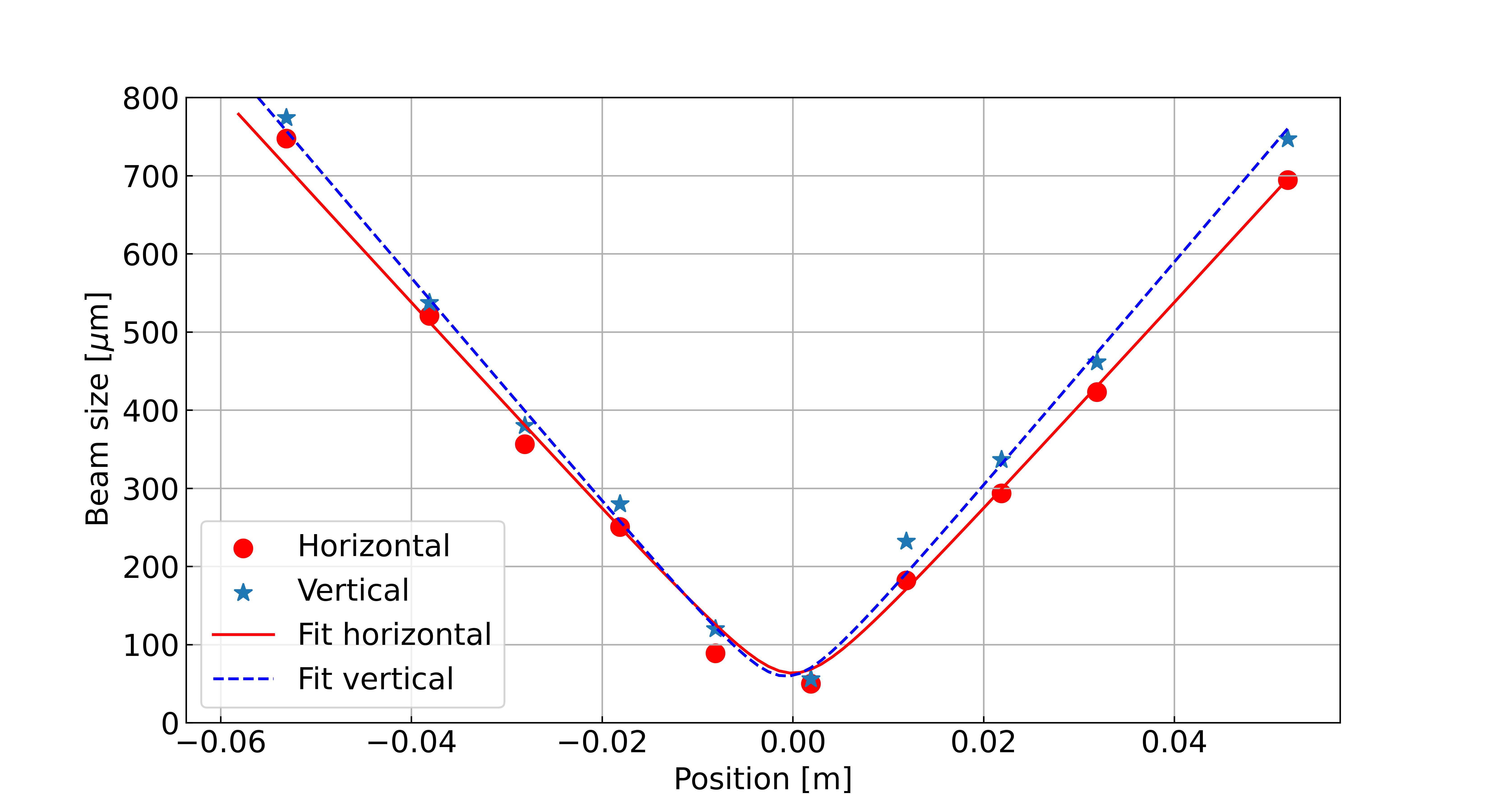}
\caption{Beam profile of the auxiliary laser. Red circle and blue star are beam radii of horizontal and vertical directions, respectively. Red solid and blue dashed lines are fitted curves.}
\label{Fig:beamprofile}
\end{figure}

The beam profile of the downconverted $2128\unit{nm}$ auxiliary light was measured by the BP209-IR2, a scanning-slit beam profiler by Thorlabs.
During this measurement, the $1064\unit{nm}$ pump light was properly filtered out by using a harmonic beam separator (HBS) and a silicon window.
Fig. \ref{Fig:beamprofile} shows the measured beam profile, which maintains the Gaussian beam shape of the $1064\unit{nm}$ pump light.
Also, the beam's $M^2$ factor is estimated to be about $1.1$, which is adequate to be used as an auxiliary beam source in a GWD.

The conversion efficiency of our OPO is about $10\%$.
This low conversion efficiency results from the under-coupled OPO cavity, and could be improved by adjusting the cavity parameters as shown in the previous work \cite{Darsow-Fromm2020}.
It is, however, not limiting the demonstration of the ALS scheme.

The downconverted auxiliary beam is phase-modulated by an electro-optic modulator (EOM) for PDH locking to the main cavity.
In our setup, the auxiliary laser is injected from the output side of the main cavity in the same manner as the previous works \cite{Izumi2012, Mullavey2012}.
The input beam powers are $6.5\unit{mW}$ and $2.5\unit{mW}$ for $1064\unit{nm}$ and $2128\unit{nm}$ beams, respectively.
At the input or output port of the cavity, the main and the auxiliary beams are combined or separated by using an HBS.
The auxiliary PDH control signal is fed back to the cavity length through a PZT on the output mirror.
PDH locking for the main laser is also enabled by actuating the PZT.
The transmitted main and auxiliary laser powers are monitored by a silicon PD and an extended InGaAs PD, respectively.
The main cavity has finesses of $\sim250$ for the $1064\unit{nm}$ main laser and $\sim30$ for the $2128\unit{nm}$ laser.
The parameters of the main cavity are listed on Table \ref{tab:MainCav}.


\begin{figure}[ht!]
\centering
\includegraphics[width=\linewidth]{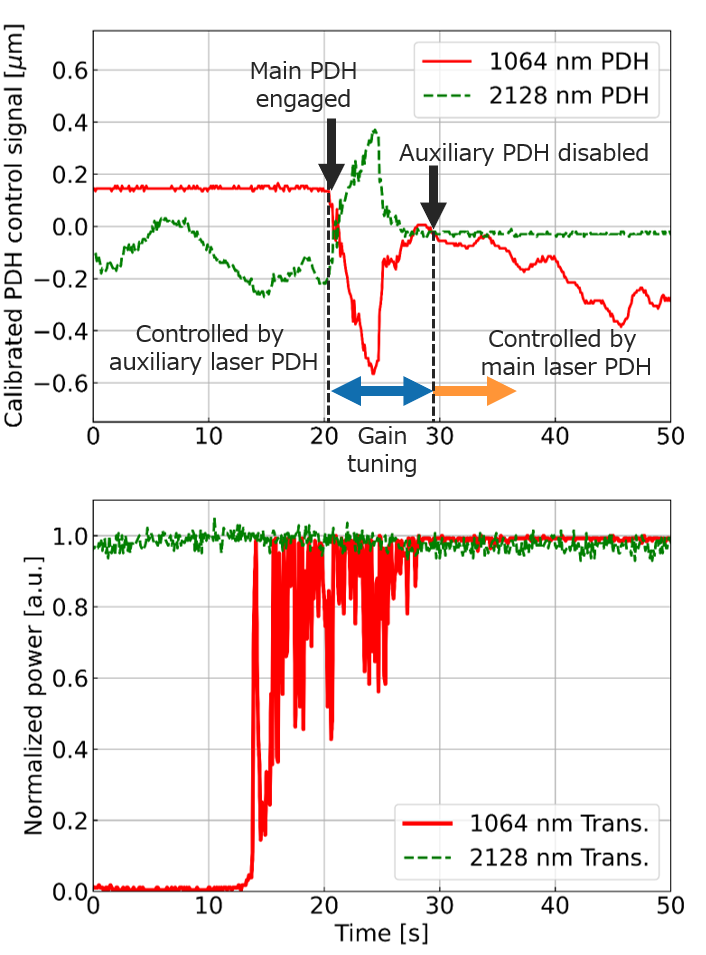}
\caption{Time series of handing over the cavity length control from the auxiliary PDH to the main PDH.
Top figure shows the PDH control signals of the main (solid red line) and auxiliary (dashed green line) lasers which are used to suppress disturbances.
These signals were calibrated by multiplying the PZT controller gain ($30\unit{V/V}$), and the PZT efficiency ($1.7\times10^{-1}\,\mu\mathrm{m/V}$) to the measured voltage. Where the control signals are constant, the corresponding loop gain is zero.
Bottom figure is the normalized transmitted powers of the main (solid red line) and auxiliary (dashed green line) lasers.}
\label{Fig:HandOff}
\end{figure}

Fig. \ref{Fig:HandOff} shows the time series data of demonstrating our ALS system.
We monitored the main and auxiliary PDH control signals and transmitted beam powers by a Tektronix MDO34 4-channel oscilloscope.
The cavity length was initially stabilized by the auxiliary laser PDH signal.
On the other hand, the main laser was kept off resonance because the two NPRO lasers had a frequency offset controlled by the DFD.
By adjusting the frequency offset, the main laser was brought to the resonant condition.
Once it reached resonance, the frequency offset tuning was stopped and the main PDH control was turned on and its servo gain was increased.
At the same time, the auxiliary PDH control gain was reduced to hand over the cavity length control to the main laser.
After fine tuning the main PDH gain, the auxiliary PDH control was disabled and the cavity was controlled by the main laser PDH.
At this stage, the auxiliary PDH control signal was constant since its control was disabled.
As shown in Fig. \ref{Fig:HandOff}, this process was successfully achieved without losing the cavity lock, i.e., the lock acquisition was achieved with $2128\unit{nm}$ downconverted auxiliary laser.
We repeated this process several times on different days, and succeeded in handing over every time, suggesting there was no intrinsic reliability issue in the proposed ALS scheme.

In addition to the demonstration of lock acquisition, we evaluated the performance of the ALS system.
Here the goal was set for the displacement of the cavity in root mean square (RMS) to be below the full width half-maximum (FWHM) linewidth, in the same manner as the previous work \cite{Mullavey2012, Izumi2012}.
The FWHM linewidth can be given by
\begin{equation}
    L_{\mathrm{FWHM}} = \frac{\lambda}{2\mathscr{F}}\unit{[m]},
    \label{eq.FWHM}
\end{equation}
where $\lambda$ is the laser wavelength and $\mathscr{F}$ is the cavity finesse.
From Eq. (\ref{eq.FWHM}), the FWHM cavity linewidth for the main $1064\unit{nm}$ light becomes
\begin{equation}
    L_{\mathrm{FWHM,\,Main}} = 1.2\unit{nm} \times \left(\frac{450}{\mathscr{F}}\right).
\end{equation}
Here we normalized the cavity finesse by that of the arm cavity used in the Advanced LIGO detector.
For the case of the Advanced LIGO arm cavity, the FWHM linewidth is $1.2\unit{nm}$.
For our tabletop setup, the FWHM linewidth is about $2.2\unit{nm}$.

\begin{figure}[ht!]
\centering
\includegraphics[width=\linewidth]{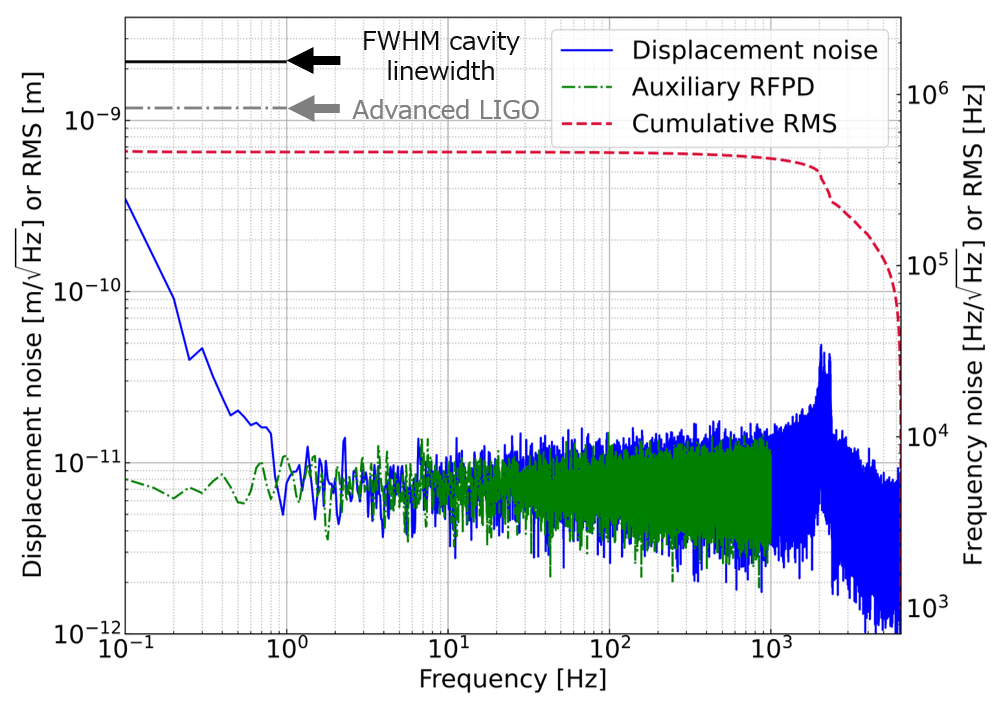}
\caption{Residual cavity displacement noise spectrum (blue solid line), RMS (red dashed line), and dark noise of the auxiliary RFPD (green dash-dotted line).
Black solid and grey dash-dotted lines are FWHM cavity linewidth of our setup and Advanced LIGO arm cavity, respectively.
The relationship between the displacement noise (left axis) and the frequency noise (right axis) can be given by Eq. (\ref{eq:cavity}). 
The features around $2\unit{kHz}$ are due to the mechanical resonances of the cavity mirror mount.}
\label{fig:noise}
\end{figure}

In order to evaluate the displacement noise of the ALS system, we used the main PDH signal as an out-of-loop sensor as shown in Fig. \ref{Fig:setup}.
Fig. \ref{fig:noise} shows the measured displacement noise spectrum and its RMS.
The right axis of this figure is the frequency noise, $\delta\nu$, which can be derived by
\begin{equation}
    \frac{\delta\nu}{\nu} = \frac{\delta L}{L},
    \label{eq:cavity}
\end{equation}
where $\nu$ is the laser frequency, $\delta L$ is the cavity displacement noise, and $L$ is the cavity length.
The measured displacement RMS was about $0.7\unit{nm}$ at $0.1\unit{Hz}$.
Therefore, the cavity displacement noise is successfully suppressed below its FWHM linewidth by the downconverted ALS system, as well as below that of Advanced LIGO's arm cavity.
Note that the requirement in terms of laser frequency will become more stringent for longer arm cavities - we will discuss this scaling below.

The performance above $\sim1\unit{Hz}$ is limited by an auxiliary PDH related noise, especially a dark noise introduced by a commercially available broadband RFPD used for $2128\unit{nm}$ PDH lock.
This noise was measured by monitoring the output from the auxiliary RFPD without injecting the light.
Below $\sim1\unit{Hz}$, air turbulence inside the cavity is considered as one of the limiting noise sources.
One possible coupling mechanism of air turbulence is fluctuations of the refractive index of air which has the wavelength dependence \cite{Quoc2009}.
Also, air turbulence can couple to the displacement noise through the beam jitter.
More detailed investigation on the impact of air turbulence can be found in Appendix C of Ref. \cite{Mullavey_thesis}.

The results above show that the main laser can be kept off resonance with cavity displacement noise less than its FWHM line width. The main laser is no longer accidentally resonated in the cavity, which would hinder the robust lock acquisition in GWDs.
By tuning the frequency offset between two lasers, the main laser can be brought to the resonance, handing over the PDH control, and demonstrating the feasibility of an ALS system based on downconverted $2128\unit{nm}$ light.

Finally, we discuss the noise scaling of this ALS system as we move from the $40\unit{cm}$ test cavity to km-scale GWD arm cavities. While the displacement requirement does not change, the requirement in terms of frequency stability scales inversely with the arm length.
For Advanced LIGO's $4\unit{km}$ arm length this requires a $10^4$ times more stringent frequency stability. 

As shown in Fig. \ref{fig:noise}, the measured noise above $1\unit{Hz}$ with our setup is about $7\times10^3\unit{Hz/{\sqrt{Hz}}}$ in terms of frequency noise, and it is limited by the auxiliary PDH related noise, specifically the dark noise in the RFPD.
Since the cavity PDH response is also proportional to the arm cavity length, the contribution of the auxiliary PDH noise will be reduced by the same factor of $10^4$, to about $0.7\unit{Hz/\sqrt{Hz}}$, and this contribution to the RMS noise will remain within the cavity line width. The RFPD dark noise can be further improved by switching to a custom resonant diode.

As for the scaling of air turbulence and vibrations, we fist note that the cavity itself will be under ultra-high vacuum (about $10^{-9}\unit{Torr}$), and all mirror vibrations are common to the two optical fields, so we can ignore noise from the cavity.
On the other hand, air turbulence and table vibrations on the in-air path of the ALS system (i.e. beat note sensing) do matter, but they will couple via Doppler effect, i.e. 
\begin{equation}
    \frac{\delta\nu}{\nu} = \frac{\delta v}{c},
\end{equation}
where $\delta v$ is the effective mirror motion (either true motion or apparent motion due to air turbulence), and $c$ is the speed of light.
In order to keep this noise level below $1\unit{Hz/\sqrt{Hz}}$, the effective mirror motion needs to be below about $1\,(\mu\mathrm{m/s)/\sqrt{Hz}}$, corresponding to $0.17 \,\mu\mathrm{m/\sqrt{Hz}} \times 1\unit{Hz}/f$.
This is the same requirement as for the current Advanced LIGO ALS system. 


The frequency discriminator noise also does not scale with the cavity length, and will be limited by the noise in the delay line (about $1\unit{Hz/\sqrt{Hz}}$).
This noise, however, can be improved by employing a phase-locked loop (PLL) \cite{Wolaver1991}.
The auxiliary laser is tightly locked to the stabilized main laser by the PLL, resulting in the laser frequency discriminator noise well below the required level \cite{Staley2014, Akutsu_2020}.

{The last nontrivial noise source which does not scale with the arm cavity length is the intrinsic frequency noise of the OPO. 
OPOs operated below threshold are routinely used for squeezed light generation in 
 GWD, and have demonstrated RMS phase noise of about $1\unit{mrad}$ \cite{Oelker2016}.
 For a flat frequency noise spectrum and a lower cutoff frequency of $1\unit{Hz}$ this corresponds to a frequency noise of $1\unit{mHz/\sqrt{Hz}}$, which is low enough for the ALS system. 
 The OPO frequency noise can be directly measured by using a Mach-Zehnder interferometer as shown in Ref. \cite{Yeaton-Massey2012}, which set a bound on the frequency noise in SHG resonators.}

As a future work, we are planning to replace one of the amorphous coated mirrors in the main dichroic cavity with an AlGaAs mirror which mimics the end test mass (ETM) of Advanced LIGO.
Such AlGaAs mirrors require optimized layers not only to have high reflectivities at $1064\unit{nm}$ and $2128\unit{nm}$, but also to reduce thermo-optic noise \cite{Chalermsongsak2016}.
Combining with a phase camera, our setup will serve as a test bed to characterize such dichroic AlGaAs mirrors under high beam intensity \cite{Muniz2021}.

It should be mentioned that one of the advantages of using $2128\unit{nm}$ light is the ability to separate the two auxiliary beams of the two perpendicular arm cavities.
The Advanced LIGO ALS system separates the green optical axis from the infrared optical axis by wedges on the beamsplitter, input test masses, and compensation plates \cite{Staley2014, Aasi2015}.
This is enabled by the difference in indices of refraction of fused silica for $532\unit{nm}$ and $1064\unit{nm}$ light \cite{Malitson1965}.
Since the $2128\unit{nm}$ light has similar difference in the refractive index compared to the $1064\unit{nm}$ light, wedges of the same size can be used for separating $2128\unit{nm}$ light from two arm cavities.
Thus, our proposed ALS system will be able to utilize the existing beam separation scheme.


As a summary, we have demonstrated a novel ALS system with frequency downconverted light.
While crystalline AlGaAs coatings can reduce the coating thermal noise by a factor of at least $5$ compared to that of current Advanced LIGO, they are not compatible with the existing green ALS system.
The ALS scheme presented in this letter will allow the use of crystalline AlGaAs coatings.
Consequently, the detection rate of binary neutron star mergers can be enhanced by a factor of about 40, leading to dramatically improved scientific impact.









\begin{backmatter}
\bmsection{Funding}
This work was supported with funding from the National Science Foundation grant PHY-2207640.

\bmsection{Acknowledgments}
The authors thank Peter Fritschel, Yutaro Enomoto, and the LIGO Scientific Collaboration Optics Working Group for useful discussions and feedback.
This paper has LIGO Document number LIGO-P2400168.

\bmsection{Disclosures} The authors declare no conflicts of interest.

\bmsection{Data availability} Data underlying the results presented in this paper are not publicly available at this time but may be obtained from the authors upon reasonable request.


\end{backmatter}

\bibliography{ref}

\begin{thebibliography}{10}
\newcommand{\enquote}[1]{``#1''}

\bibitem{Abbott2016}
B.~P. Abbott, R.~Abbott, T.~D. Abbott, \emph{et~al.}, \enquote{Observation of gravitational waves from a binary black hole merger,} {\protect\JournalTitle{Phys. Rev. Lett.}} \textbf{116}, 061102 (2016).

\bibitem{Abbott2021}
R.~Abbott, T.~D. Abbott, S.~Abraham, \emph{et~al.}, \enquote{Tests of general relativity with binary black holes from the second ligo-virgo gravitational-wave transient catalog,} {\protect\JournalTitle{Phys. Rev. D}} \textbf{103}, 122002 (2021).

\bibitem{Aasi2015}
J.~Aasi \emph{et~al.}, \enquote{Advanced {LIGO},} {\protect\JournalTitle{Classical and Quantum Gravity}} \textbf{32}, 074001 (2015).

\bibitem{Mullavey2012}
A.~J. Mullavey, B.~J.~J. Slagmolen, J.~Miller, \emph{et~al.}, \enquote{Arm-length stabilisation for interferometric gravitational-wave detectors using frequency-doubled auxiliary lasers,} {\protect\JournalTitle{Opt. Express}} \textbf{20}, 81--89 (2012).

\bibitem{Izumi2012}
K.~Izumi, K.~Arai, B.~Barr, \emph{et~al.}, \enquote{Multicolor cavity metrology,} {\protect\JournalTitle{J. Opt. Soc. Am. A}} \textbf{29}, 2092--2103 (2012).

\bibitem{Staley2014}
A.~Staley \emph{et~al.}, \enquote{{Achieving resonance in the Advanced {LIGO} gravitational-wave interferometer},} {\protect\JournalTitle{Classical and Quantum Gravity}} \textbf{31}, 245010 (2014).

\bibitem{Akutsu_2020}
T.~Akutsu \emph{et~al.}, \enquote{An arm length stabilization system for {KAGRA} and future gravitational-wave detectors,} {\protect\JournalTitle{Classical and Quantum Gravity}} \textbf{37}, 035004 (2020).

\bibitem{Acernese2023}
F.~Acernese, M.~Agathos, A.~Ain, \emph{et~al.}, \enquote{The advanced virgo+ status,} {\protect\JournalTitle{Journal of Physics: Conference Series}} \textbf{2429}, 012039 (2023).

\bibitem{Gras2018}
S.~Gras and M.~Evans, \enquote{Direct measurement of coating thermal noise in optical resonators,} {\protect\JournalTitle{Phys. Rev. D}} \textbf{98}, 122001 (2018).

\bibitem{Buikema2020}
A.~Buikema \emph{et~al.}, \enquote{{Sensitivity and performance of the Advanced LIGO detectors in the third observing run},} {\protect\JournalTitle{Phys. Rev. D}} \textbf{102}, 062003 (2020).

\bibitem{LSC}
{LIGO Scientific Collaboration}, \enquote{{LSC Instrument Science White Paper},} \url{https://dcc.ligo.org/LIGO-T2300411/public} (2023).

\bibitem{Cole2023}
G.~D. Cole, S.~W. Ballmer, G.~Billingsley, \emph{et~al.}, \enquote{{Substrate-transferred GaAs/AlGaAs crystalline coatings for gravitational-wave detectors},} {\protect\JournalTitle{Applied Physics Letters}} \textbf{122}, 110502 (2023).

\bibitem{Adachi1985}
S.~Adachi, \enquote{{GaAs}, {AlAs}, and $\mathrm{Al}_{x}\mathrm{Ga}_{1-x}\mathrm{As}$: Material parameters for use in research and device applications,} {\protect\JournalTitle{Journal of Applied Physics}} \textbf{58}, R1--R29 (1985).

\bibitem{Steinlechner2021}
J.~Steinlechner and I.~W. Martin, \enquote{How can amorphous silicon improve current gravitational-wave detectors?} {\protect\JournalTitle{Phys. Rev. D}} \textbf{103}, 042001 (2021).

\bibitem{PDH}
R.~W.~P. Drever, J.~L. Hall, F.~V. Kowalski, \emph{et~al.}, \enquote{Laser phase and frequency stabilization using an optical resonator,} {\protect\JournalTitle{Applied Physics B: Lasers and Optics}} \textbf{31}, 97--105 (1983).

\bibitem{Darsow-Fromm2020}
C.~Darsow-Fromm, M.~Schr\"{o}der, J.~Gurs, \emph{et~al.}, \enquote{Highly efficient generation of coherent light at 2128 nm via degenerate optical-parametric oscillation,} {\protect\JournalTitle{Opt. Lett.}} \textbf{45}, 6194--6197 (2020).

\bibitem{Quoc2009}
T.~B. Quoc, M.~Ishige, Y.~Ohkubo, and M.~Aketagawa, \enquote{Measurement of air-refractive-index fluctuation from laser frequency shift with uncertainty of order $10^{-9}$,} {\protect\JournalTitle{Measurement Science and Technology}} \textbf{20}, 125302 (2009).

\bibitem{Mullavey_thesis}
A.~Mullavey, \enquote{Arm length stabilisation for advanced gravitational wave detectors,} Ph.D. thesis, Australian National University (2012).

\bibitem{Wolaver1991}
D.~Wolaver, \emph{Phase-locked Loop Circuit Design}, Prentice Hall advanced reference series (Prentice Hall, 1991).

\bibitem{Oelker2016}
E.~Oelker, G.~Mansell, M.~Tse, \emph{et~al.}, \enquote{Ultra-low phase noise squeezed vacuum source for gravitational wave detectors,} {\protect\JournalTitle{Optica}} \textbf{3}, 682--685 (2016).

\bibitem{Yeaton-Massey2012}
D.~Yeaton-Massey and R.~X. Adhikari, \enquote{A new bound on excess frequency noise in second harmonic generation in ppktp at the 10{\textminus}19 level,} {\protect\JournalTitle{Opt. Express}} \textbf{20}, 21019--21024 (2012).

\bibitem{Chalermsongsak2016}
T.~Chalermsongsak, E.~D. Hall, G.~D. Cole, \emph{et~al.}, \enquote{Coherent cancellation of photothermal noise in {GaAs}/{$\mathrm{Al_{0.92}Ga_{0.08}As}$} bragg mirrors,} {\protect\JournalTitle{Metrologia}} \textbf{53}, 860--868 (2016).

\bibitem{Muniz2021}
E.~Mu\~niz, V.~Srivastava, S.~Vidyant, and S.~W. Ballmer, \enquote{High frame-rate phase camera for high-resolution wavefront sensing in gravitational-wave detectors,} {\protect\JournalTitle{Phys. Rev. D}} \textbf{104}, 042002 (2021).

\bibitem{Malitson1965}
I.~H. Malitson, \enquote{Interspecimen comparison of the refractive index of fused silica,} {\protect\JournalTitle{J. Opt. Soc. Am.}} \textbf{55}, 1205--1209 (1965).

\end{thebibliography}




\end{document}